\documentclass[a4paper]{jpconf}
\usepackage[utf8]{inputenc}
\usepackage{graphicx}
\usepackage{verbatim}
\usepackage{comment}
\usepackage{amssymb}
\usepackage{amsmath}
\usepackage{ mathrsfs }
\usepackage{enumerate}
\usepackage{psfrag}
\usepackage{pstricks}
\usepackage{color}
\usepackage{hyperref}
\usepackage{enumitem}
\usepackage{float}
\usepackage{braket}
\usepackage{physics}
\usepackage{xcolor}

\begin{document}

\bibliographystyle{iopart-num}

\title{Ground state phase diagram of the one-dimensional Bose-Hubbard model from restricted Boltzmann machines}
\author{Kristopher McBrian}
\address{Department of Physics and Astronomy, San Jose State University, San Jose, CA 95192, USA}
\author{Giuseppe Carleo}
\address{Center for Computational Quantum Physics, Flatiron Institute, New York, NY 10010, USA}
\author{Ehsan Khatami}
\address{Department of Physics and Astronomy, San Jose State University, San Jose, CA 95192, USA}

\begin{abstract}
Motivated by recent advances in the representation of ground state wavefunctions of 
quantum many-body systems using restricted Boltzmann machines as variational ansatz,
we utilize an open-source platform for constructing such 
ansatz called NetKet to explore the extent of 
applicability of restricted Boltzmann machines to bosonic lattice models. Within NetKet, we 
design and train these machines for the one-dimensional Bose-Hubbard model through a 
Monte Carlo sampling of the Fock space. We vary parameters such as the strength of the onsite 
repulsion, the chemical potential, the system size and the maximum site occupancy and use 
converged equations of state to identify phase boundaries between the Mott insulating 
and superfluid phases. We compare the average density and the energy to results from 
exact diagonalization and map out the ground state phase diagram, which agrees qualitatively 
with previous finding obtained through conventional means.
\end{abstract}

\section{Introduction}

In recent years, artificial neural networks have been employed to study the Bose-Hubbard model~\cite{h_saito_17,h_saito_18}, 
which describes the many-body systems of interacting bosonic particles on lattices~\cite{m_fisher_89}. 
In Ref.~\cite{h_saito_17}, the author approaches the problem using feedforward artificial neural networks 
where the inputs are Fock states (see below) and the output layer, consisting of two nodes, provides 
the $Re$ and $Im$ parts of the log of the wavefunction. The networks are trained following conventional 
supervised learning algorithms for feedforward networks with the cost function taken to be the ground 
state energy. Ref.~\cite{h_saito_18} expands on this idea and examines the role multiple hidden layers 
or convolutional layers play in the accuracy of the output and the efficiency of the algorithm. 

Restricted Boltzmann machines (RBMs) have also emerged as useful 
tools among artificial neural networks for providing a superior ansatz for representing quantum wave 
functions~\cite{Carleo2016,d_deng_16,g_torlai_17a,j_chen_18,y_huang_17}, 
or mimicking thermodynamics of classical and quantum many-body 
systems~\cite{Torlai2016,g_torlai_17,y_nomura_17}. For example, in Ref.~\cite{Carleo2016}, 
it was shown that RBMs lead to lower variational ground state energy for classical and quantum 
magnetic models than the state-of-the-art ansatz. This in turn motivated several other studies 
in which the RBMs were used to represent topological~\cite{d_deng_16,g_torlai_17a} or chiral~\cite{y_huang_17} states,
draw equivalences between RBMs and tensor network states~\cite{j_chen_18}, and to accelerate 
Monte Carlo simulations~\cite{l_huang_17}, to name a few.

Here, we use RBMs to represent ground state wavefunctions of the one-dimensional
Bose-Hubbard model and study the extent to which its equilibrium properties can be captured in this 
architecture for a range of model parameters. It has been shown that {\em multi-valued} RBMs, such as the 
one we have used here, can represent
a wide class of many-body entangled states efficiently~\cite{s_lu_18}, and to the best of our knowledge, our results are the first to demonstrate that. 
We perform the training using Monte Carlo 
sampling with the same goal of minimizing the ground state energy as in Refs.~\cite{h_saito_17,h_saito_18}.
We employ NetKet~\cite{netket}, an open-source platform for solving quantum many-body problems using artificial intelligence. 
We show that RBMs can be trained to represent the ground state 
of the bosonic system with a good degree of accuracy.  We find that the 
energy and density both in the Mott insulating and superfluid phases rapidly converge to final values during the training. By 
calculating equations of state for different values of the interaction strength in the grand canonical ensemble, 
we map out the phase diagram of the 
model and show that it agrees well with one obtained in the past directly from quantum Monte Carlo simulations 
on larger systems.\\

\section{The Bose-Hubbard Model}

The Bose-Hubbard Hamiltonian is expressed as
\begin{gather}
H=-t\sum_{\left < ij \right >}b_i^\dagger b_j + \frac{U}{2}\sum_i n_i(n_i - 1) - \mu\sum_i n_i,
\end{gather}
where $b_i^\dagger$ and $b_j$ are the bosonic creation and annihilation operators, respectively, 
$\left < ij \right >$ indicates that $i$ and $j$ are nearest neighbors, $t$ is the corresponding hopping 
amplitude, which we set to unity to serve as the unit of energy throughout the paper, 
$U$ is the strength of the onsite repulsion, $n_i=b_i^\dagger b_i$ is the site occupation operator, 
and $\mu$ is the chemical potential. In this study, we consider the periodic one-dimensional (1D) geometry.

For large enough interactions, the ground state of the system is the interaction-driven Mott insulating 
phase while it remains a superfluid for smaller interactions. This was shown in a seminal paper by Fisher 
et al.~\cite{m_fisher_89} where the universality class of the model was explored and a mean-field solution 
was presented. Later, the model received much attention in the 1990's, mostly confirming numerically 
the transition in different 
dimensions~\cite{g_batrouni_90,r_scalettar_91,p_niyaz_91,d_rokhsar_91,w_krauth_91b,w_krauth_91,w_krauth_92,a_kampf_93,c_bruder_93,a_vanotterlo_95,n_prokofev_98}. 
The Mott insulator (MI) to superfluid (SF) transition was also observed in experiments with cold atoms 
in optical lattices~\cite{d_jaksch_98}.

\section{Restricted Boltzmann Machine}

\subsection{The Architecture}

\begin{figure}[t]
\centering
\includegraphics[width=12pc]{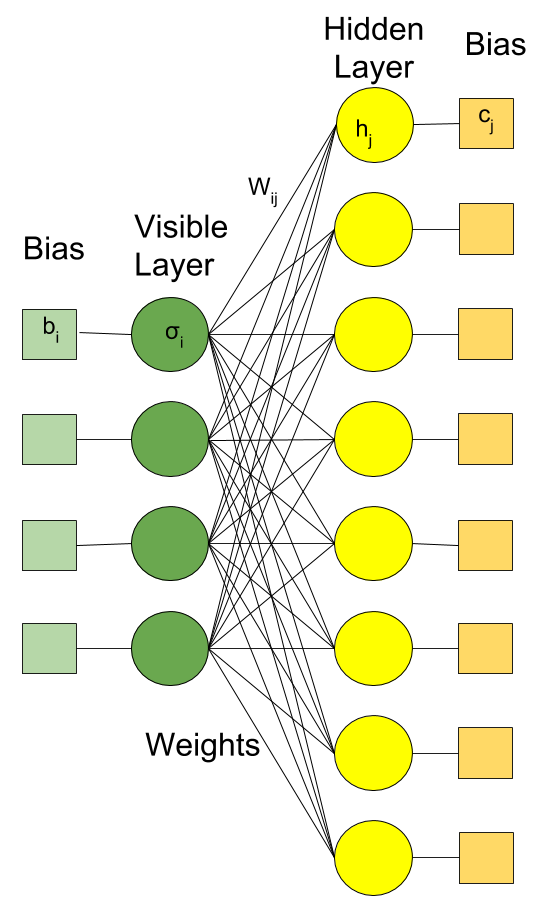}\hspace{2pc}%
\begin{minipage}[b]{19pc}\caption{\label{label} Sample architecture of the RBM used in this study. 
We choose visible layers of different sizes for $L=4, 5$, $6$, and $7$, which also represent the system sizes and twice as
many nodes in the hidden layer as in the input layer.  Both the input nodes and hidden nodes take 
binary values; $\sigma_i$ and $h_i$, respectively. Every site occupation $n_j = \{0,1,...,N_{max}$\} is
converted to $N_{max}+1$ binary digits, all 0's except the $n_j$th digit, which is 1. Therefore, the number
of input nodes is $(N_{max}+1)\times L$. Each node has a bias value (shown as $b_i$ or $c_i$) and nodes
are fully connected between the two layers with weights $w_{ij}$ associated with each connection. The weights
and biases are determined through a stochastic optimization process that involves local Metropolis moves between 
Fock states where the expectation value of the Bose-Hubbard Hamiltonian is minimized.}\label{fig:rbm}
\end{minipage}
\end{figure}

The RBM is a two-layer fully-connected artificial neural network. The architecture is shown in Fig.~\ref{fig:rbm}. 
It has one visible layer, and one hidden layer. Each layer consists of several nodes with an external bias 
parameter associated with them ($b_i$ for the visible and $c_i$ for the hidden layer). Each node in a layer 
is connected to all the nodes in the other layer and a weight parameter ($W_{ij}$) is associated to every 
connection. 

The RBM lends itself well to representing ground state wavefunctions of the Bose-Hubbard model because a probability distribution of possible Fock states of the system can then be calculated directly from the RBM parameters after training. This is similar to representing thermodynamics of a many-body system using an RBM where a probability distribution of microstates can be extracted~\cite{g_torlai_17}.

Within NetKet, we define a wave function ansatz for the $q=0$ momentum sector using a RBM with spin-1/2 hidden 
nodes and permutation symmetry. 
The input layer can be visualized as a two-dimensional $L\times (N_{max}+1)$ array of binary numbers, where 
each row represents a site, and a 1 in column number $n$ indicates that there are $n$ bosons at that site. 
This so-called ``one-hot" encoding of site occupations is not unique. For example, an alternative would be to use $L$ non-binary
input nodes whose values can vary from $0$ to $N_{max}$ instead. One advantage of the one-hot representation 
over the latter is that the number of network parameters is much larger and therefore, the training will presumably be more refined.
We choose $L$ and $N_{max}$ values as large as 7 and 5, respectively, and a hidden layer with twice as many nodes as in 
the visible layer.

\subsection{The Training}

We write the ground state wave function as a linear combination of the Fock state basis 
\begin{equation}
\ket{\psi} = \sum_n \psi_n \ket{n},
\end{equation}
where $\ket{n}=\ket{n_1,n_2,\dots,n_L}$ are the particle number Fock states with site occupancies $n_i\in [0,N_{max}]$. 
The goal is to create a RBM that can stochastically represent $\ket{\psi}$ by producing the correct probability, 
$|\psi_n|^2$ for every input Fock state $\ket{n}$ through the input layer. For the Bose-Hubbard model, the coefficients 
$\psi_n$ are known to be real and positive and are written for the RBM as 

\begin{equation}
\psi_n = \sum_{\{h\}} e^{\sum_i b_i \sigma_i + \sum_j c_j h_j + \sum_{ij} W_{ij} \sigma_i h_j}.
\end{equation}
where $\sigma_i$ and $h_i$ denote the input node and hidden node binary values, respectively.

The hidden degrees of freedom can be integrated out exactly for the simple architecture shown in Fig.~\ref{fig:rbm}, 
\begin{eqnarray}
\sum_{\{h\}} e^{\sum_i b_i \sigma_i + \sum_j c_j h_j + \sum_{ij} W_{ij}\sigma_i h_j}
&=& e^{\sum_i b_i \sigma_i} \sum_{\{h\}}\prod_j^{N_h} e^{c_j h_j + \sum_i W_{ij}\sigma_i h_j}\nonumber\\
 &=& e^{\sum_i b_i \sigma_i} \prod_j^{N_h} \sum_{h_j=0,1}e^{c_j h_j + \sum_i W_{ij}\sigma_i h_j}\nonumber\\
 &=& e^{\sum_i b_i \sigma_i} \prod_j^{N_h} (1 + e^{c_j + \sum_i W_{ij}\sigma_i})\nonumber\\
 &=& e^{\sum_i b_i \sigma_i} \prod_j^{N_h} e^{\log(1 + e^{c_j + \sum_i W_{ij}\sigma_i})},
\end{eqnarray}
which
results in~\cite{Carleo2016,Torlai2016}
\begin{equation}
\psi_n = e^{\sum_i b_i \sigma_i + \sum_j \log \left (1+e^{c_j + \sum_i W_{ij} \sigma_i} \right )}.
\end{equation}

The goal in the training of the RBM is to optimize its weights and biases in order to arrive at a probability 
distribution for $\psi_n$ so that $\ket{\psi}$ represents the  ground state. This is accomplished through 
the minimization of the expectation value of the Hamiltonian,
\begin{equation}
E=\frac{\expval{\psi|H|\psi}}{\braket{\psi}}.
\label{eq:E}
\end{equation}

NetKet adopts a Metropolis algorithm for importance sampling of $\ket{n}$'s during which $E$ is minimized 
also using stochastic techniques. Since we are working in the grand canonical ensemble, we choose local 
moves in which a site is picked at random and its occupation is proposed to change to a value in the range
$[0,N_{max}]$. The move is accepted with the probability min$[1,|\psi_{n'}|^2/|\psi_n|^2]$, where $n'$ and $n$
are the new and old states, respectively. This guarantees that after the training is completed, the probability 
distribution is $P_n=|\psi_n|^2/\sum_m |\psi_m|^2 =|\psi_n|^2/\braket{\psi}$. The minimization of energy is 
done using the Stochastic Reconfiguration algorithm~\cite{f_becca_17} with AdaMax optimizer in NetKet and
the default parameters. Our sampler performs 1000 sweeps before sampling the probability distribution
and updating the RBM parameters. More details about the method 
and the training parameters can be found in Ref.~\cite{netket}.

\section{Results}

\begin{figure}[t]
\centering
\includegraphics[width=0.7\textwidth]{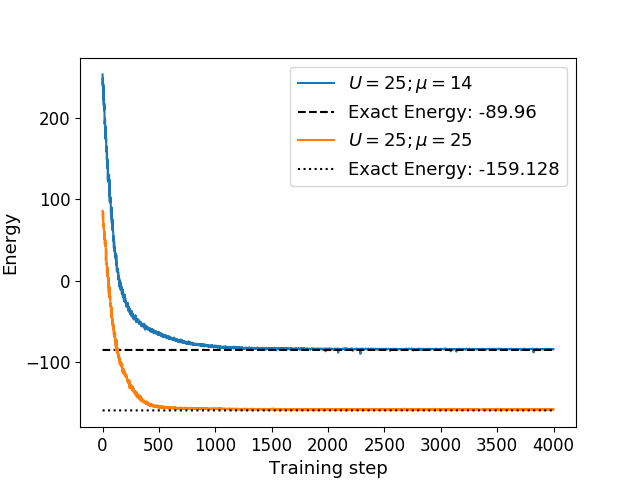}%\hspace{2pc}%
\begin{minipage}[b]{10pc}\caption{\label{fig:energy}Training a RBM using stochastic reconfiguration aims to find the ground state of the system.  
This is done by minimizing the expectation value of the energy, which is plotted above as a function of training step.  
This convergence is an indication that the system has entered the ground state of the Hamiltonian. Here, $L=6$ and $N_{max} = 5$.}
\end{minipage}
\end{figure}

\begin{figure}[t]
\centering
\includegraphics[width=0.7\textwidth]{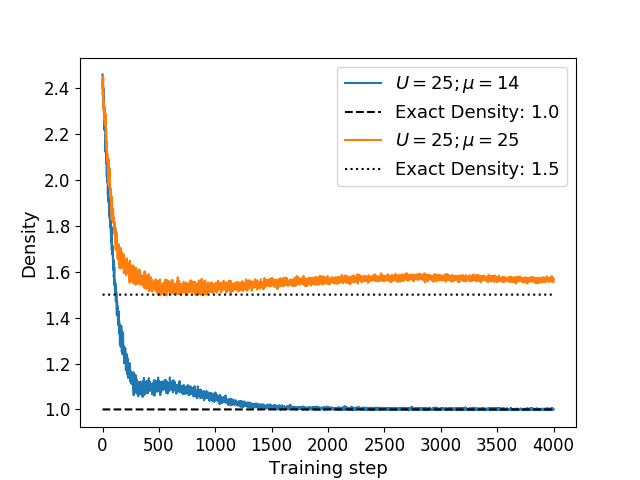}%\hspace{2pc}%
\begin{minipage}[b]{10pc}\caption{\label{fig:density} Same as in Fig.~\ref{fig:energy} but for the evolution of the average density over training step.}
\end{minipage}
\end{figure}

In Figs.~\ref{fig:energy} and \ref{fig:density}, we show the average energy, $E$, and the average density, $\rho$, 
as the training of our RBM progresses for two sets of values for the interaction strength and the chemical potential.
The system with $\mu/U=0.56$ (blue lines) is a Mott insulator with a density of one particle per site and the system with 
$\mu/U=1.00$ (orange lines) is a superfluid for the system size we have considered ($L=6$). Here, $N_{max} = 5$ 
is much larger than the average density. We observe that the energies and densities quickly converge to their final 
values regardless of whether the system is in the MI or SF phase. However, we find that the average density has 
larger fluctuations over the training steps inside the SF phase while it is pinned to unity when the system is in the 
first Mott lobe.

For comparison, we also show in Figs.~\ref{fig:energy} and \ref{fig:density} results from exact diagonalization
as horizontal dashed and dotted lines. We find that the error in both the energy and the density is larger in the 
superfluid phase. The fluctuations in the density seem to extend to very long training times and point to the 
limitations of the method and the fact that more complex networks or better optimization may be needed to capture the physics in 
the superfluid phase.

\begin{figure}[t]
  \centering
   \includegraphics[width=0.7\textwidth]{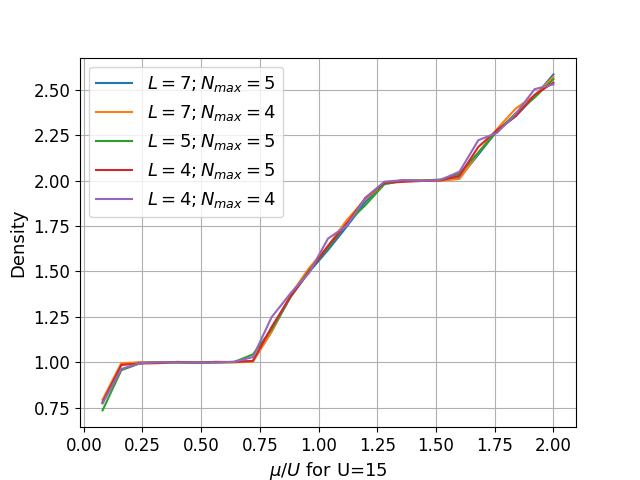}
  \caption{Equation of state for a fixed interaction strength of $U=15$, different system sizes and different limits for the 
 boson site occupation numbers. The density, $\rho$, does not show significant variations by changing the system size or 
 the maximum occupation numbers as long as $N_{max}>\rho$.}\label{eq:eos}
\end{figure}

Taking long-time values of the converged density as we vary the chemical potential at a fixed interaction strength, 
we obtain the equation of state. Figure~\ref{eq:eos} shows this property for $U/t=15$ and several different combinations
of $L$ and $N_{max}$. We find the expected behavior~\cite{m_fisher_89,g_batrouni_90} where the density is pinned 
at integer numbers of bosons per site as we cross multiple SF and MI phase boundaries. Two density plateaus corresponding to the 
first and second Mott lobes are clearly visible in this plot. Increasing the system size from $4$ to $7$ seems to 
lead to smoother curves, signaling better convergence during the training, but no appreciable change in the onsets of the 
Mott regions. Moreover, the results do not change significantly by increasing $N_{max}$ from $4$ to $5$. 
These observations do not mean that our results are not suffering from finite size effects. Much larger clusters have been 
shown to give significantly different results in conventional treatments of the model~\cite{g_batrouni_90,e_satoshi_12}.

\begin{figure}[t]
\centering
\includegraphics[width=0.7\textwidth]{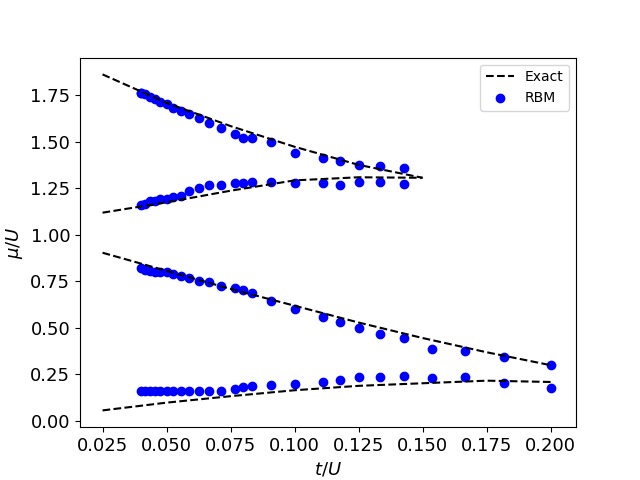}
\caption{\label{fig:pd} Ground state phase diagram of the Bose-Hubbard 
model for $L=6$ and $N_{max} = 5$ in the $t/U$-$\mu/U$ space from the RBM. The first two Mott lobes 
for $\rho=1$ and $\rho=2$ are clearly captured in the strong-coupling region where $t/U\ll 1$. We use a 
tolerance of $\pm 0.05$ for the density around integer values to determine the departure from the MI phases. 
Dashed lines are exact results for $L=128$ from Ref.~\cite{e_satoshi_12}.}
\end{figure}

By training different RBMs as we sweep the chemical potential at increasing values of the interaction strength, 
we are able to map out the ground state phase diagram of the model in the interaction-chemical potential space.
The findings are summarized in Fig.~\ref{fig:pd}. We extend the results up to the second Mott lobe. They qualitatively
agree with those obtained using Monte Carlo algorithms and much larger system sizes (see e.g. Ref.~\cite{e_satoshi_12}).
As with other techniques, the accurate locating of the phase boundary near the tip of the Mott lobes are the most
challenging due to the Kosterlitz-Thouless nature of the transition~\cite{j_carrasquilla_13}.

\section{Summary}

Using the open-source package NetKet, recently developed for the study of quantum many-body systems through 
machine learning algorithms, we design and train RBMs to represent the ground state wavefunction of the 
1D Bose-Hubbard model on small chains. We demonstrate the degree of applicability of the RBM ansatz in various regions 
of the parameter space as MI or SF phases set in. We show that the optimization of free parameters of the 
RBMs during training in order to minimize the ground state energy leads to physical properties, including 
expectation values of the Hamiltonian and a phase diagram, that agree with known exact results from conventional
treatments of the model.

\section{Acknowledgements}
KM acknowledges support from Undergraduate Research Grants at San Jose State University. 
EK acknowledges support from the National Science Foundation (NSF) under Grant No. DMR-1609560.

\newpage

\section{References}

\providecommand{\newblock}{}

\end{document}